\documentclass[aps,reprint,prl,floatfix,amsmath,amssymb,preprintnumbers,amsfonts]{revtex4-2}
\usepackage[pdftex]{graphicx}
\usepackage{color}
\usepackage{dcolumn}
\usepackage{comment}

\begin{document}
\title{
Neural-Network-Assisted Boltzmann Approach for Dilute Microswimmer Suspensions
}
\author{Haruki Hayano}
\thanks{hayano@iis.u-tokyo.ac.jp}
\author{Akira Furukawa}
\thanks{furu@iis.u-tokyo.ac.jp}
\affiliation{Institute of Industrial Science,
The University of Tokyo, Meguro-ku, Tokyo 153-8505, Japan}
\author{Kang Kim}
\thanks{kk@cheng.es.osaka-u.ac.jp}
\affiliation{Division of Chemical Engineering, Department of Materials Engineering Science, Graduate School of Engineering Science,
The University of Osaka, Toyonaka, Osaka 560-8531, Japan}
\date{\today}
\begin{abstract}
We introduce a neural-network--assisted Boltzmann framework that learns the binary-collision map of microswimmers directly from data and uses it to evaluate collision integrals efficiently. Using a representative model swimmer, the learned map quantitatively predicts translational and rotational diffusivities and enables a linear-stability analysis of isotropy against polar ordering in dilute suspensions. The resulting predictions closely match direct simulations. The present framework is agnostic to active matter models and broadly applicable: once two-body collision data are obtained---either from simulations or experiments---the same surrogate can be used to evaluate kinetic transport across dilute conditions where binary collisions dominate. Because the workflow relies only on pre- and post-collision statistics, the present approach provides a general data-driven route linking particle-scale interactions to macroscopic transport and collective behavior in active suspensions.
\end{abstract}

\maketitle
Suspensions of microswimmers, including bacteria, algae, and synthetic active colloids, exhibit dynamical behaviors absent in passive colloids, such as anomalous rheology, mesoscale turbulence, and motility-induced transport \cite{Ramaswamy2010,Marchetti2013,Lauga2016,Alert2022,Baconnier2025,Sokolov2009,Lopez2015,Wensink2012,Wu2000,Mino2011,Lin2011,Thiffeault2010,Morozov2014}.
These phenomena arise from the interplay of self-propulsion, hydrodynamic interactions (HIs), steric and chemical interactions, fluctuations, and boundary effects or external fields.

Interparticle/intercellular interactions govern both local dynamics and emergent collective states.
In mesoscopic descriptions such as Smoluchowski kinetics and continuum hydrodynamics, their effects are often incorporated into effective collision kernels and transport coefficients \cite{Saintillan2008,Wensink2012,Saintillan2018,Baskaran2008,Baskaran2009}.
Obtaining reliable values for these quantities requires resolving the elementary collisional processes, particularly those controlled by near-field HIs and steric contact.
In dilute suspensions, binary collisions dominate, and a Boltzmann-type kinetic equation that treats them explicitly provides a natural coarse-graining route \cite{Aranson2005,Bertin2006,Bertin2009,Peshkov2012,Hanke2013,Weber2013,Thuroff2013,Peshkov2014,Suzuki2015,Oyama2017,Grossmann2020}.
However, near-field coupling makes the two-body scattering map highly nonlinear and sensitive to swimmer geometry and propulsion type (pusher, puller, neutral) \cite{Alexander2008,Drescher2011}.
While detailed 2-body solutions exist for specific model swimmers such as squirmers \cite{Ishikawa2006_2,Papavassiliou2017,Gotze2010}, extending those results to arbitrary swimmer designs remains challenging.

In this Letter, we present a neural-network-assisted Boltzmann approach that learns the binary-collision map of microswimmers from pre- and post-collisional data generated by direct hydrodynamic simulations, and uses it to evaluate collision integrals efficiently.
For any chosen swimmer model, providing its collision statistics enables \emph{quantitative} predictions of rotational and translational diffusivities and of whether the isotropic state is stable or polar order can emerge.
As a representative demonstration, our predictions agree with many-body hydrodynamic simulations in the dilute regime $(\phi \lesssim 0.04)$. 
Because our approach relies only on collision statistics, the same workflow can be applied directly to other swimmer classes once their binary-collision data are available.

\begin{figure*}[tbh]
\includegraphics[width=0.96\textwidth]{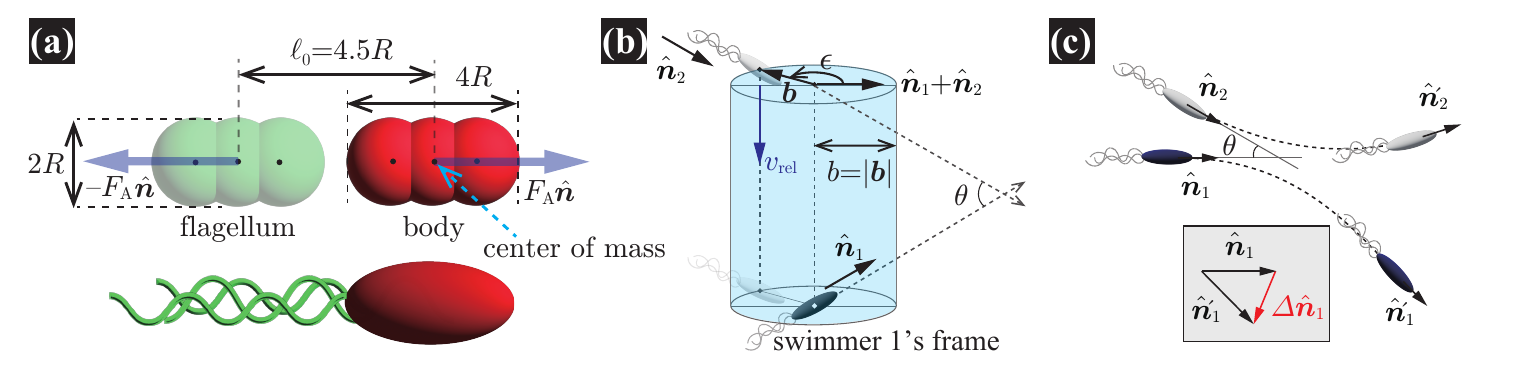}
\caption{(color online)
(a) Schematic of our model microswimmer resembling {\it E. coli}. The body is treated as rigid, whereas the flagellum is modeled as a massless ``phantom'' particle. Both the body and the flagellum are represented by three overlapping spheres of radius $R$. Further details of the swimmer model and its construction are provided in Ref.\cite{Hayano2022}.
(b) Pre-collisional state of two microswimmers specified by the relative angle $\theta$, impact parameter $b$, and azimuthal angle $\epsilon$. The vector $\mbox{\boldmath $b$}$ connecting the centers of the two swimmers at the closest approach in the absence of any interactions is perpendicular to the relative velocity: $\mbox{\boldmath $b$} \perp \mbox{\boldmath $v$}_{\rm rel}$. See Supplementary Material for the details of the initial configuration in our simulations.
(c) Schematic of a binary collision between two microswimmers with relative angle $\theta$. The pre-collisional orientations $\hat{\mbox{\boldmath$n$}}_1$ and $\hat{\mbox{\boldmath$n$}}_2$ change to the post-collisional orientations $\hat{\mbox{\boldmath$n$}}'_1$ and $\hat{\mbox{\boldmath$n$}}'_2$ after the collision. We denote the orientation changes as $\Delta\hat{\mbox{\boldmath$n$}}_1 = \hat{\mbox{\boldmath$n$}}'_1 - \hat{\mbox{\boldmath$n$}}_1$ and $\Delta\hat{\mbox{\boldmath$n$}}_2 = \hat{\mbox{\boldmath$n$}}'_2 - \hat{\mbox{\boldmath$n$}}_2$.
}
\label{fig:modelswimmer}
\end{figure*}

{\it Model swimmers and simulation method}---
For the preparation of dataset, we conduct direct hydrodynamic simulations of a binary collision of two model microswimmers.
Our model swimmer used in this study is exactly the same as that used in our previous work \cite{Hayano2022}, which is a simplified model of a rod-like swimmer with a flagellum such as {\it E. coli}.
The model microswimmer, schematically shown in Fig. \ref{fig:modelswimmer}(a), is composed of body and flagellum parts.
The body part is treated as a rigid body, while the flagellum part is regarded as a massless “phantom” particle simply following the body’s motions.
The body part is exerted a self-propulsion force $F_{\rm A} \hat{\mbox{\boldmath$n$}}$ while the flagellum part exerts the force $-F_{\rm A} \hat{\mbox{\boldmath$n$}}$ directly on the solvent fluid.
Here, $\hat{\mbox{\boldmath$n$}}$ is the unit vector of the orientation of the swimmer.
Full details of the model swimmer can be found in our previous work \cite{Hayano2022}.
The hydrodynamic interactions (HIs) between swimmers are incorporated using the Smoothed Profile Method \cite{Nakayama2005,Molina2016,Yamamoto2021}, which is one of the mesoscopic simulation techniques that can treat HIs in a computationally efficient manner.

{\it Binary collision of microswimmers}---
A binary collision between two swimmers with orientations
$\hat{\mbox{\boldmath $n$}}_1$ and $\hat{\mbox{\boldmath $n$}}_2$
is specified by three geometric parameters $(\theta, b, \epsilon)$, as illustrated in Fig. \ref{fig:modelswimmer}(b).
The three parameters are defined as follows:
(i) the relative angle is defined as $\theta \equiv \arccos(\hat{\mbox{\boldmath $n$}}_1 \cdot \hat{\mbox{\boldmath $n$}}_2)$, $(0<\theta\leq\pi)$.
(ii) Let $\mbox{\boldmath $b$}$ be the vector connecting the centers of the two swimmers at the closest approach in the absence of any interactions, whose direction is perpendicular to the relative velocity $\mbox{\boldmath $v$}_{\rm rel} = v_{\rm s}(\hat{\mbox{\boldmath $n$}}_1 - \hat{\mbox{\boldmath $n$}}_2)$, where $v_{\rm s}$ is the swimming speed of an isolated swimmer.
The impact parameter is defined as $b \equiv |\mbox{\boldmath $b$}|$, $(0\leq b < \infty)$.
(iii) The azimuthal angle $\epsilon$, $(-\pi < \epsilon \leq \pi)$ is the angle between $\mbox{\boldmath $b$}$ and $\hat{\mbox{\boldmath $n$}}_1+\hat{\mbox{\boldmath $n$}}_2$.
For identical axisymmetric swimmers, particle-exchange and axial symmetries allow the range limited to $0 \leq \epsilon \leq \pi$.

Based on $(\theta,b,\epsilon)$, we generated a dataset of 2,048 simulated collisions storing the post-collisional orientations $(\hat{\mbox{\boldmath $n$}}'_1,\hat{\mbox{\boldmath $n$}}'_2)$, schematically illustrated in Fig. \ref{fig:modelswimmer}(c).
The parameters ranges were chosen as follows: $\pi/16 \le \theta \le \pi$, because for small relative angles the orientational change is sufficiently small to be neglect and introducing a lower cutoff in $\theta$ does not affect the predictions; $0 \le b \le 2\ell_{0}$, since the orientational change is negligible for $b>2\ell_{0}$ in our simulations (refer to the Supplemental Material, SM), where $\ell_{0}$ is the swimmer length, and $0 \le \epsilon \le \pi$.

{\it Training and evaluation of neural network}---
We train a neural network (NN) to predict the post-collisional state of a binary collision from its pre-collisional configuration.
The NN is a fully connected feed-forward model with three hidden layers, each consisting of 384 neurons with ReLU activation.
The input layer contains four variables representing the pre-collisional state: $\cos\theta$, $\sin\theta$, $b\cos\epsilon$, and $b\sin\epsilon$.
The output layer contains six variables corresponding to the post-collisional orientation vectors of the two swimmers,
$\hat{\mbox{\boldmath$n$}}'_1 = (\hat{n}'_{1,x}, \hat{n}'_{1,y}, \hat{n}'_{1,z})$ and
$\hat{\mbox{\boldmath$n$}}'_2 = (\hat{n}'_{2,x}, \hat{n}'_{2,y}, \hat{n}'_{2,z})$. 

The network is trained using the Adam optimizer with a learning rate of $8 \times 10^{-4}$ and a batch size of 16.
The loss function is defined as the mean squared error between the predicted and simulated post-collisional orientation vectors.
Training is performed for 100 epochs with a 4:1 split between training and validation data, and cross-validation is used to suppress overfitting.
The relative error and mean angular error of the trained NN are evaluated on the test dataset (10\% of the total dataset and not included in the training data) and found to be $4.7 \%$ and $0.048\ {\rm rad}(=2.7^{\circ})$, respectively.
All computations are implemented in Python using TensorFlow with the Keras API \cite{tensorflow2015-whitepaper}.

\begin{figure}[tbh]
\includegraphics[width=0.48\textwidth]{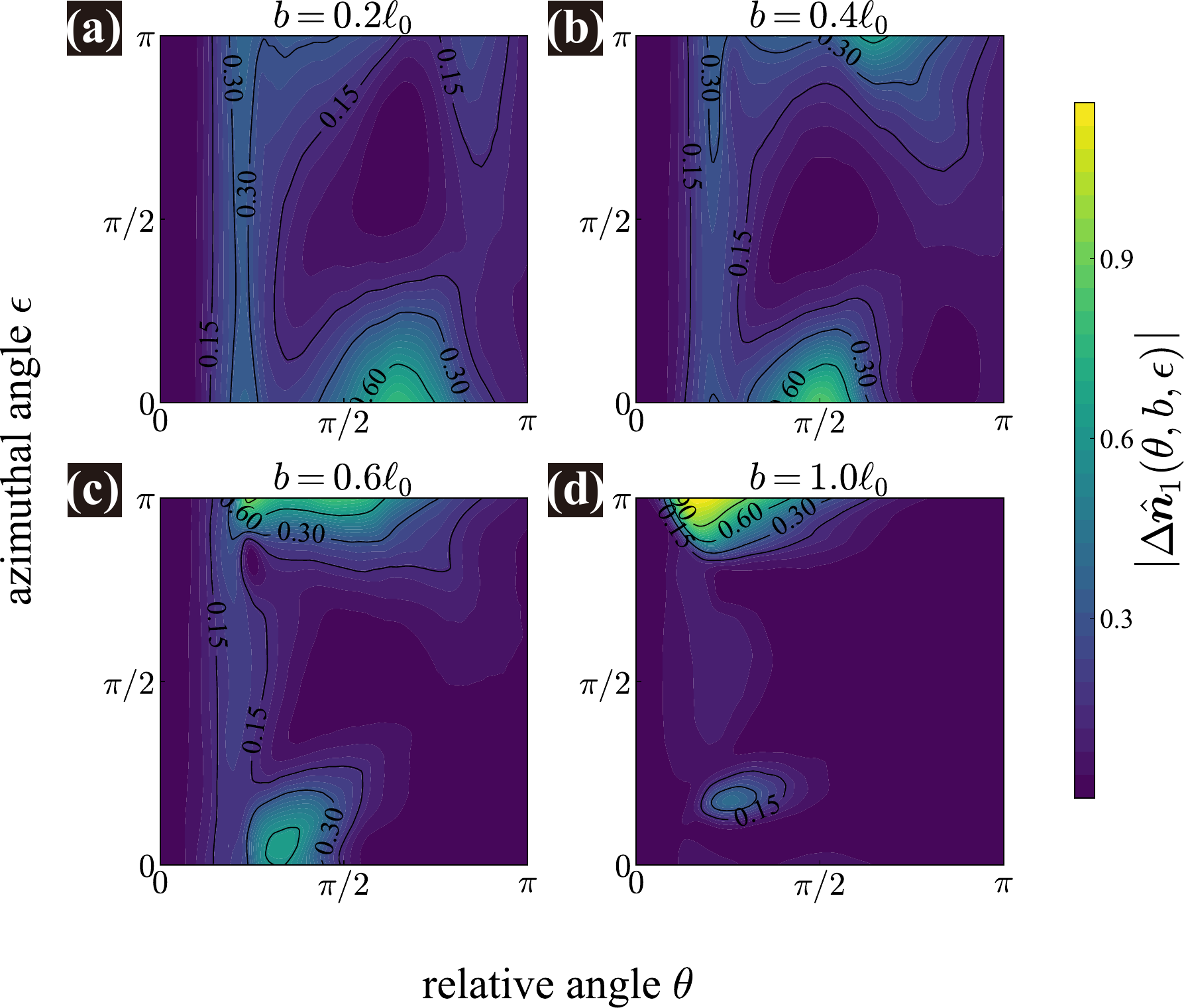}
\caption{(color online)
  Predicted orientational change of the swimmer 1, $|\Delta \hat{\mbox{\boldmath$n$}}_1| = |\hat{\mbox{\boldmath$n$}}'_1 - \hat{\mbox{\boldmath$n$}}_1|$, as a function of the relative angle $\theta$, azimuthal angle $\epsilon$ and impact parameter $b$.
  The color indicates the magnitude of the orientational change.
}
\label{fig:NNoutput}
\end{figure}

Figure~\ref{fig:NNoutput} shows the predicted orientational change of swimmer 1, $|\Delta \hat{\mbox{\boldmath$n$}}_1| = |\hat{\mbox{\boldmath$n$}}'_1 - \hat{\mbox{\boldmath$n$}}_1|$, by the collision with pre-collisional parameters $(\theta, b, \epsilon)$.
The results indicate a strong dependence on the impact parameter $b$.
For smaller $b$, swimmer 1 undergoes large orientational changes over a wide range of initial conditions $(\theta,\epsilon)$.
When the two swimmers pass each other at close proximity, near-field HIs are strong enough to substantially modulate their trajectories during the encounter. 
In particular, for initial values $\theta \cong \pi/4$ and $(\theta,\epsilon)=(\pi/2,0)$, the extensile flow generated in their front regions effectively acts as a repulsive/repelling interactions, producing pronounced reorientation.
In contrast, for larger $b$, the trajectories are only weakly perturbed over the passing time, and significant reorientation occurs only within a narrow range of $(\theta,\epsilon)$ that corresponds to direct head-to-tail collisions ---specifically, between the head of swimmer 2 and the tail of swimmer 1. 
In our model, the flagellum (tail) is treated as a massless phantom particle, while direct repulsive (excluded-volume) interactions between swimmers are retained.
As a result, when a swimmer collides with another's flagellum---located far from the latter's center of mass---the large lever-arm effect produces a significant torque, which in turn causes a larger reorientation of the swimming direction.
Note that, for most $b$ values, the configuration $(\theta,\epsilon)=(\pi/2,\pi/2)$ produces minimal reorientation because this parameter corresponds to a side-by-side configuration where the swimmers weakly interact hydrodynamically.

{\it Evaluation of diffusivity by kinetic theory with a trained NN}---
We evaluate self-diffusion in dilute microswimmer suspensions by isolating the
binary-collision contribution via a kinetic-theory calculation in which the post-collision orientation change is supplied by a trained neural network (NN), and we validate the prediction against direct many-body hydrodynamic simulations.

The translational diffusion coefficient $D_{\rm t}$ is defined by
\begin{equation}
  D_{\rm t}=\lim_{t\to\infty}\frac{\langle|\Delta\mbox{\boldmath$R$}(t)|^2\rangle}{6t},
\label{eq:def_Dt}
\end{equation}
where $\langle\cdots\rangle$ denotes an average over swimmers and time origins, and
$\Delta\mbox{\boldmath$R$}(t)$ is the displacement of the swimmer's center of mass (explicitly defined in SM).
Analogously, the rotational diffusion coefficient $D_{\rm r}$ is
\begin{equation}
  D_{\rm r}=\lim_{t\to\infty}\frac{\langle|\Delta\mbox{\boldmath$\Theta$}(t)|^2\rangle}{4t},
\label{eq:def_Dr}
\end{equation}
with $\Delta\mbox{\boldmath$\Theta$}(t)$ the unbounded angular displacement.

To focus on the effect of binary collisions, we evaluate the Boltzmann collision
integral with the integrand (single-collision orientation change) provided by the NN.
The binary-collision contribution to rotational diffusion is
\begin{equation}
  D^{\rm (bc)}_{\rm r}
  = \lim_{t\to 0}\frac{\langle|\Delta\hat{\mbox{\boldmath$n$}}(t)|^2\rangle}{4t}
  = \frac{1}{4}\int {\rm d}\Gamma(\theta,b,\epsilon)\,
    |\Delta\hat{\mbox{\boldmath$n$}}(\theta,b,\epsilon)|^2,
\label{eq:DrBC1}
\end{equation}
where ${\rm d}\Gamma=\rho\,v_{\rm rel}\,P(\theta)\,{\rm d}\theta\,{\rm d}\sigma$ is the differential collision rate. 
Here $\rho$ is the number density, $v_{\rm rel}=2v_{\rm s}\sin(\theta/2)$ the relative speed, $P(\theta)=\sin\theta/2$ the isotropic distribution of relative angles, and ${\rm d}\sigma=b\,{\rm d}b\,{\rm d}\epsilon$ the differential cross-section \footnote{Equations \eqref{eq:DrBC1} gives the short-time slope of the mean-square orientation change produced by a single (uncorrelated) binary collision. If the swimming direction decorrelates sufficiently between collisions, as in dilute suspensions, this short-time slope coincides with the long-time rotational diffusivity defined in Eq.~\eqref{eq:def_Dr}.}.
Explicitly,
\begin{eqnarray}
  D^{\rm (bc)}_{\rm r}
  &=& \frac{\rho v_{\rm s}}{4}\int_0^{b_{\rm max}}{\rm d}b\int_0^{\pi}{\rm d}\theta \int_{-\pi}^{\pi}{\rm d}\epsilon \nonumber \\
      &&b\,\sin\theta\,\sin(\tfrac{\theta}{2})
      \left|\Delta\hat{\mbox{\boldmath$n$}}(\theta,b,\epsilon)\right|^2 .
\label{eq:DrBC2}
\end{eqnarray}
Similar expressions are found in the literatures~\cite{Lin2011,Mino2011,Thiffeault2010,Morozov2014}.

\begin{figure}[tbh]
\includegraphics[width=0.48\textwidth]{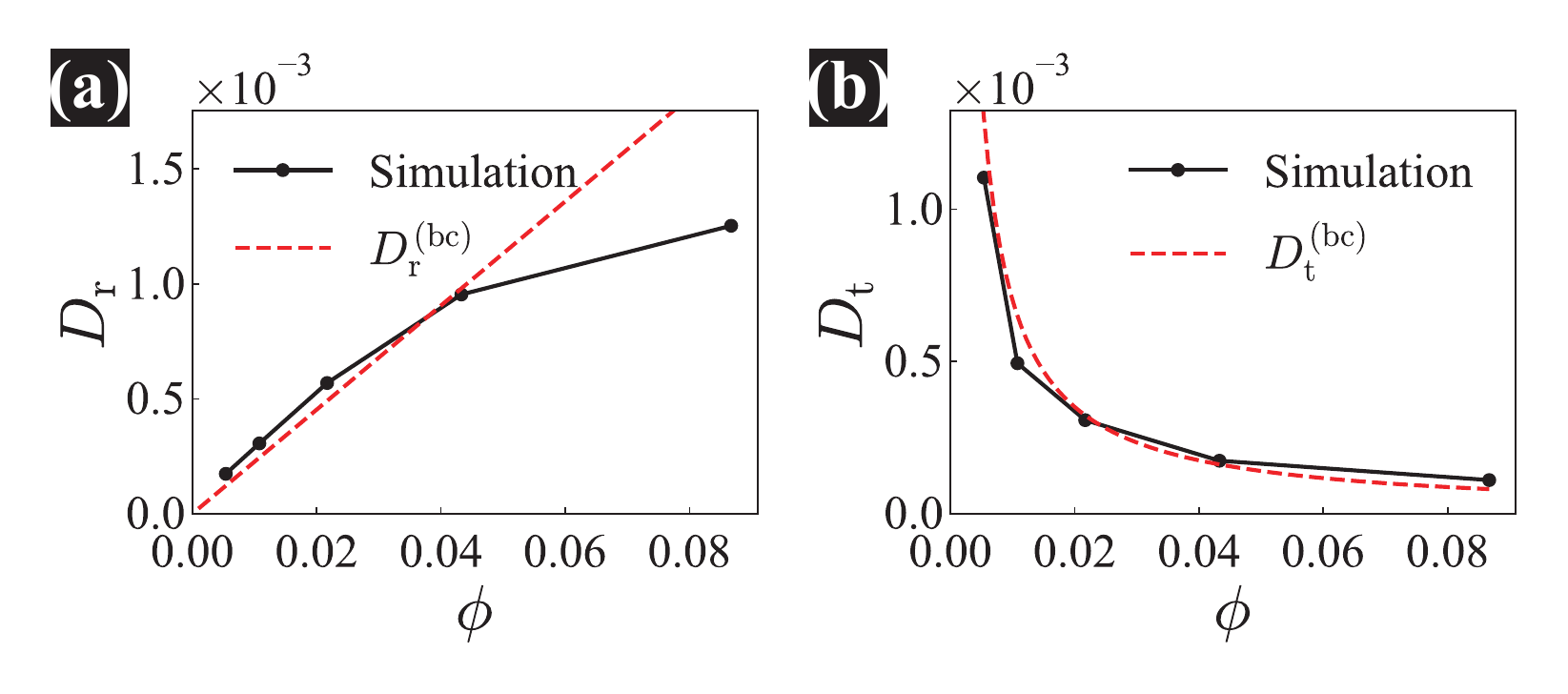}
\caption{(color online)
(a) Rotational diffusion coefficient $D_{\rm r}$ from many-body simulations (solid line)
and the binary-collision prediction $D_{\rm r}^{\rm (bc)}$ from Eq.~\eqref{eq:DrBC2}
(dashed line) versus swimmer volume fraction $\phi$.
(b) Translational diffusion coefficient $D_{\rm t}$ from simulations (solid line) and
$D_{\rm t}^{\rm (bc)}$ from Eq.~\eqref{eq:DtBC} (dashed line) versus $\phi$.}
\label{fig:diffusivity}
\end{figure}

Figure~\ref{fig:diffusivity}(a) compares $D_{\rm r}$ from many-body simulations with the kinetic-theory prediction Eq.~\eqref{eq:DrBC2} using the trained NN varying the volume fraction of the body parts $\phi$.
In evaluating Eq.~\eqref{eq:DrBC2}, we set $b_{\rm max}=2\ell_0$ and use uniform quadrature with 100 points per variable.
As shown in Fig.~\ref{fig:diffusivity}(a), $D_{\rm r}$ and $D_{\rm r}^{\rm (bc)}$
agree for $\phi\lesssim0.04$, indicating that binary collisions dominate rotational
diffusion in the dilute regime.

Assuming a constant swimming speed $v_{\rm s}$, the translational diffusion coefficient follows from a Green--Kubo relation as
\begin{equation}
  D_{\rm t}^{\rm (bc)}
  = \frac{1}{3}\int_0^{\infty}\!\langle\mbox{\boldmath$v$}_{\rm s}(t)\!\cdot\!\mbox{\boldmath$v$}_{\rm s}(0)\rangle\,{\rm d}t
  = \frac{v_{\rm s}^2}{6 D_{\rm r}^{\rm (bc)}}.
\label{eq:DtBC}
\end{equation}
Note that this expression assumes that the swimming direction decorrelates solely via binary collisions instantaneously, and is essentially equivalent to that of a run-and-tumble particle \cite{Saintillan2018}.
Figure~\ref{fig:diffusivity}(b) shows $D_{\rm t}^{\rm (bc)}$ is almost consistent with $D_{\rm t}$ over the range examined \footnote{the agreement for $\phi \gtrsim 0.04$ may be coincidental. Although a significant deviation appears in $D_{\rm r}$ for $\phi \gtrsim 0.04$, this discrepancy is not reflected in $D_{\rm t}$.
The deviation in $D_{\rm r}$ is attributable to crowding: increasing $\phi$ slows rotational dynamics. At the same time, crowding also reduces the actual $v_{\rm s}$, whereas the evaluation of $D_{\rm t}^{\rm (bc)}$ assumes a constant dilute-limit speed. In practice, both $D_{\rm r}$ and $v_{\rm s}$ decrease with $\phi$; these two effects combine to make $D_{\rm t}$ appear consistent with $D_{\rm t}^{\rm (bc)}$ even for $\phi \gtrsim 0.04$, where the deviation in $D_{\rm r}$ is already evident.}.

{\it Linear stability of the isotropic state}---
We assess the linear stability of the spatially homogeneous isotropic state within a Boltzmann description of binary collisions \cite{Aranson2005,Bertin2006,Hanke2013,Suzuki2015}.
We neglect spatial dependence and consider only the orientational distribution
$f(\hat{\mbox{\boldmath$n$}},t)$, normalized by $\int {\rm d}\hat{\mbox{\boldmath$n$}}f=1$.
Its evolution obeys
\begin{equation}
  \partial_t f(\hat{\mbox{\boldmath$n$}},t)
  = D_{\rm r}^{({\rm th})}\,\nabla_{\hat n}^{2} f(\hat{\mbox{\boldmath$n$}},t)
  + \mathcal{C}[f],
\label{eq:Boltzmann}
\end{equation}
where $\nabla_{\hat n}$ is the gradient operator on the unit sphere, $D_{\rm r}^{({\rm th})}$ the thermal rotational diffusivity (absent in the present simulations), and $\mathcal{C}[f]$ the collision integral.

The isotropic steady state is $f_0=1/(4\pi)$. 
Considering small perturbations, we write
$f=f_0+\delta f(t)$ with
\[
  \delta f(t)=\frac{3}{4\pi}\,\mbox{\boldmath$m$}(t)\!\cdot\!\hat{\mbox{\boldmath$n$}},\qquad
  \mbox{\boldmath$m$}(t)\equiv\int\,{\rm d}\hat{\mbox{\boldmath$n$}}\,\hat{\mbox{\boldmath$n$}}\delta f(t) .
\]
Multiplying Eq.~\eqref{eq:Boltzmann} by $\hat{\mbox{\boldmath$n$}}$ and integrating over orientations yields
\begin{equation}
  \frac{{\rm d}\mbox{\boldmath$m$}(t)}{{\rm d}t}
  = \Big[-2D_{\rm r}^{({\rm th})}+\nu^{\rm (bc)}\Big]\,
    \mbox{\boldmath$m$}(t),
\end{equation}
and
\begin{equation}
  \nu^{\rm (bc)}
  = \int {\rm d}\Gamma(\theta,b,\epsilon)\;
    \big(\Delta\hat{\mbox{\boldmath $n$}}_{1}+\Delta\hat{\mbox{\boldmath $n$}}_{2}\big)
    \!\cdot\!
    \big(\hat{\mbox{\boldmath $n$}}_{1}+\hat{\mbox{\boldmath $n$}}_{2}\big).
\label{eq:nu1}
\end{equation}
Here ${\rm d}\Gamma=\rho\,v_{\rm rel}\,P(\theta)\,{\rm d}\theta\,{\rm d}\sigma$ with
$v_{\rm rel}=2v_{\rm s}\sin(\theta/2)$, $P(\theta)=\tfrac12\sin\theta$, and
${\rm d}\sigma=b\,{\rm d}b\,{\rm d}\epsilon$.
The single-collision orientation increments $\Delta\hat{\mbox{\boldmath $n$}}_{i}(\theta,b,\epsilon)$
$(i=1,2)$ are provided by the trained NN.
Note that Eq.~\eqref{eq:nu1} is the same as that derived in Ref.~\cite{Suzuki2015} for 2D systems, and see the SM for a detailed derivation of Eq.~\eqref{eq:nu1}.

The growth rate (eigenvalue) is
$\nu \equiv -2D_{\rm r}^{({\rm th})}+\nu^{\rm (bc)}$, the isotropic state is linearly stable for $\nu<0$ and unstable for $\nu >0$.
Using the trained NN we obtain $\nu^{\rm (bc)}/\Gamma_0 \simeq -0.13$, so the isotropic state remains stable even for $D_{\rm r}^{({\rm th})}=0$, where the per-particle total collision rate is $\Gamma_0=\int {\rm d}\Gamma(\theta,b,\epsilon)$.

\begin{figure}[tbh]
\includegraphics[width=0.48\textwidth]{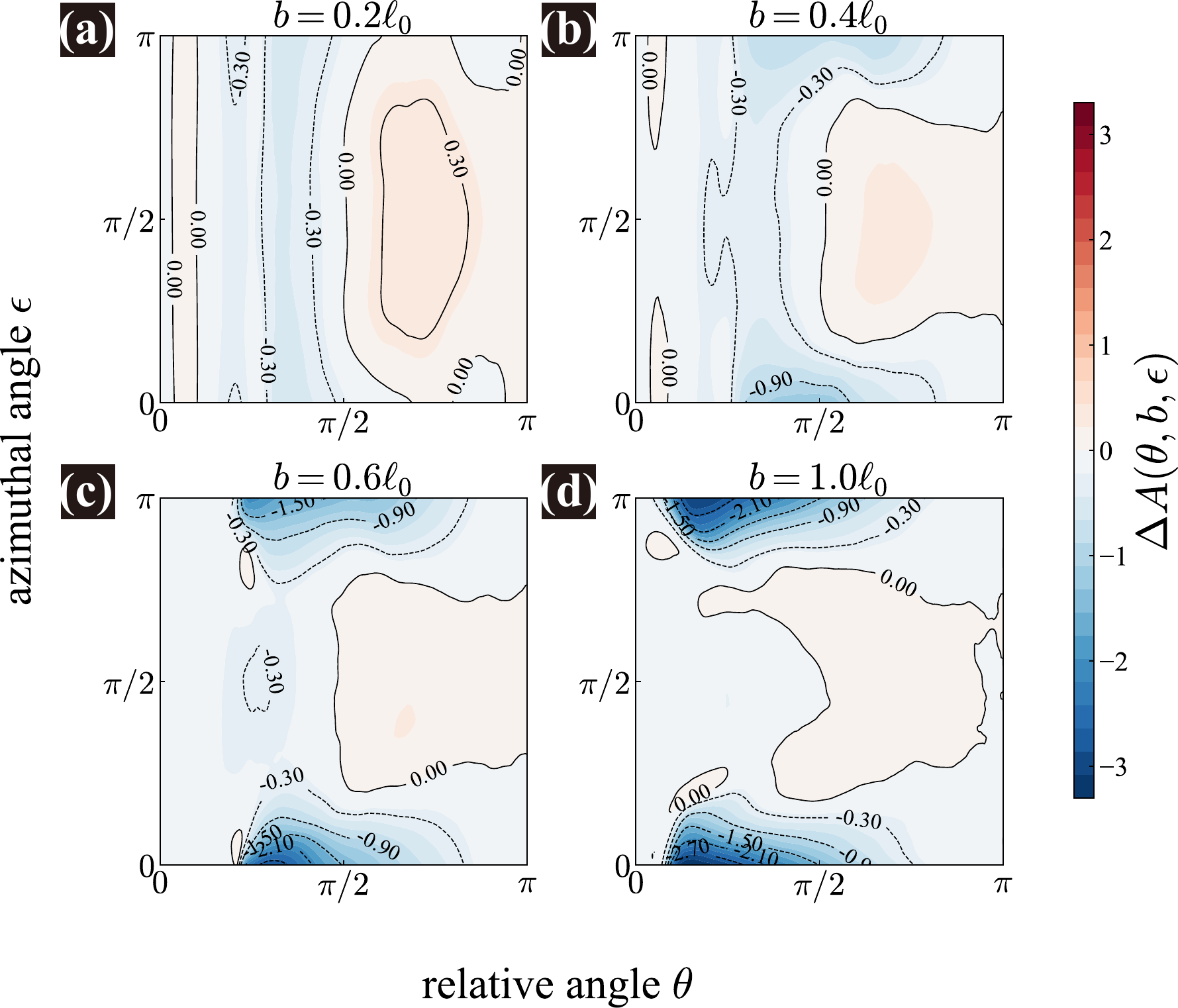}
\caption{(color online)
Alignment metric
$\Delta A \equiv
(\Delta\hat{\mbox{\boldmath $n$}}_{1}+\Delta\hat{\mbox{\boldmath $n$}}_{2})
\!\cdot\!
(\hat{\mbox{\boldmath $n$}}_{1}+\hat{\mbox{\boldmath $n$}}_{2})$
as a function of relative angle $\theta$, azimuth $\epsilon$, and impact parameter $b$.
Colors indicate the magnitude of the alignment (negative values: disalignment).}
\label{fig:mu1}
\end{figure}

To further understand the collision-induced alignment, we examine the integrand in Eq.~\eqref{eq:nu1}.
We define the alignment metric
\begin{equation}
  \Delta A(\theta,b,\epsilon)
  \equiv
  \big(\Delta\hat{\mbox{\boldmath $n$}}_{1}+\Delta\hat{\mbox{\boldmath $n$}}_{2}\big)
  \!\cdot\!
  \big(\hat{\mbox{\boldmath $n$}}_{1}+\hat{\mbox{\boldmath $n$}}_{2}\big) .
\end{equation}
Figure~\ref{fig:mu1} shows $\Delta A$ versus the collision parameters $(\theta,b,\epsilon)$.
The alignment effect is predominantly negative (disaligning), with particularly strong disalignment near $\theta\simeq \pi/4$ at small $b$, and at large $b$ for head-to-tail configurations $\epsilon\simeq 0,\pi$.
For collisions with $\theta \lesssim \pi/2$, when the swimmers are initially aligned, they tend to disalign after the collision, while for $\theta \gtrsim \pi/2$, when they are initially anti-aligned, they tend to align.
By axial symmetry $\Delta A$ is even about $\epsilon=\pi/2$, a property almost captured by the NN.

{\it Concluding remarks}---
We have introduced a neural-network-assisted Boltzmann framework that learns the
binary-collision law directly from data and evaluates collision integrals
efficiently.
For a representative microswimmer, the learned law yields quantitative predictions for rotational and translational self-diffusion in the dilute regime and matches many-body simulations for $\phi \lesssim 0.04$.
A linear analysis further shows that collisions are predominantly disaligning, stabilizing the isotropic state.
Because the surrogate model encodes only the single-encounter input-output map, it is agnostic to data provenance and applies to {\it any} self-propelled particles once collision statistics are available, whether measured experimentally or generated numerically.
This provides a compact and transferable approach from resolved pair interactions to continuum-level transport, enabling the development of kinetic theories for active matter.

{\it Acknowledgments}---
This work was supported by JSPS KAKENHI (Grants No. JP20H05619) and JSPS Core-to-Core Program ``Advanced core-to-core network for the physics of self-organizing active matter'' (Grants No. JPJSCCA20230002) and JST SPRING (Grants No. JPMJSP2108).

\bibliography{library}

\end{document}